\documentclass[aps,prx,twocolumn,superscriptaddress,showpacs]{revtex4-2}
\usepackage{braket}
\usepackage{xcolor}
\usepackage{mathtools}
\usepackage{hyperref}
\usepackage{times,xspace}
\usepackage{graphicx,graphics,color,epsfig}
\usepackage{amsmath}
\usepackage{amssymb}
\usepackage{amsfonts}
\usepackage{amsfonts}
\usepackage{epstopdf}
\usepackage{bm}
\usepackage{hyperref}
\usepackage{longtable}     % helps with long table options
\usepackage{url}           % for on-line citations
\usepackage{bm}            % special 'bold-math' package
\usepackage{color}
\usepackage{float}
\usepackage[normalem]{ulem}

\newcommand*{\bord}{\multicolumn{1}{c|}{}}

\begin{document}
	
	\title{%Superconducting pairing symmetry in Ba$_{2}$CuO$_{3+x}$ using first principle//
		%Unconventional Superconductivity in Ba$_{2}$CuO$_{3+x}$: The presence of electron pockets and pairing symmetry\\
		Doping induced singlet to triplet superconducting transition in Ba$_{2}$CuO$_{3+\delta}$ 
		%: The presence of electron pockets and pairing symmetry  
	}
	
	\author{Priyo Adhikary}
\affiliation{Center for Atomistic Modelling and Materials Design, Indian Institute of Technology Madras, Chennai, 600036, India}
\affiliation{Department of Physics, Indian Institute Of Technology Madras, Chennai, 600036, India}
\author{Mayank Gupta}
\affiliation{Center for Atomistic Modelling and Materials Design, Indian Institute of Technology Madras, Chennai, 600036, India}
\affiliation{Condensed Matter Theory and Computational Lab, Department of Physics, Indian Institute Of Technology Madras, Chennai, 600036, India}
%\author{Sashi Satpathy}  
%\affiliation{Condensed Matter Theory and Computational Lab, Department of Physics, Indian Institute Of Technology Madras, Chennai, 600036, India}
%\affiliation{Department of Physics \& Astronomy, University of Missouri, Columbia, MO 65211, USA}    
\author{B. R. K. Nanda}
\email{nandab@iitm.ac.in}
\affiliation{Center for Atomistic Modelling and Materials Design, Indian Institute of Technology Madras, Chennai, 600036, India}
\affiliation{Condensed Matter Theory and Computational Lab, Department of Physics, Indian Institute Of Technology Madras, Chennai, 600036, India}
\author{Shantanu Mukherjee}
\email{shantanu@iitm.ac.in}
\affiliation{Center for Atomistic Modelling and Materials Design, Indian Institute of Technology Madras, Chennai, 600036, India}
\affiliation{Department of Physics, Indian Institute Of Technology Madras, Chennai, 600036, India}
	\date{\today}
	
	\begin{abstract}
	In this study, we perform a numerical simulation on the recently discovered high-temperature superconductor ($T_c$= 73K) Ba$_2$CuO$_{3.2}$ \cite{lietal} while focusing on doping dependence of alternating CuO$_6$ octahedra and CuO chain-like states. Employing the multiband random-phase approximation, we compute the spin-fluctuation mediated pairing interaction, subsequently determining its pairing eigenvalues and eigenfunctions relative to oxygen-doping levels. We find that, for the certain range of hole  doping in Ba$_2$CuO$_{3+\delta}$, a singlet $d_{x^2-y^2}$-wave pairing symmetry emerges as long as we keep the doping below the critical value $x_{c}$. Interestingly upon hole doping, the dominant pairing symmetry undergoes a transition to a triplet (odd paring) type from the singlet state. This change in pairing is driven by the competition between the nesting vectors coming from the Fermi surface of $d_{z^2}$ and $d_{x^2-y^2}$ orbitals within the CuO$_6$ octahedra. This triplet state is attainable through hole doping, while supressing inter-layer self-doping effects. Furthermore, we present the density of states within the superconducting phase, offering a potential comparison with tunnelling spectra in Ba$_2$CuO$_{3+\delta}$. Our research provides novel insights into the intricate pairing symmetries in Ba$_2$CuO$_{3+\delta}$ and their underlying pairing mechanisms.
	\end{abstract}
	\maketitle
	
	%{\it Introduction.-}
	
	\section{Introduction}
	
	One of the intriguing features of cuprate superconductors is the possibility of $d$-wave pairing symmetry, that leads to a superconducting gap with nodes on the Fermi surface (FS)\cite{TsueiRMP,ScalapinoDwave,KirtleyDwave}.  However, recent experiments have pointed towards the possibility of a nodeless pairing state in certain electron and hole doped regions of the cuprate high-Tc superconductors \cite{wu,andreone,schneider,LSCO2000,LSCO2013,bi22122006,bi2212,bi2201,ccoc,YBCO}. This deviation from a $d$-wave nature of the superconducting gap suggests a more complex phase diagram of cuprates than previously considered\cite{scalapino_rmp1}. Theoretical studies have also shown that a triplet state can emerge in cuprates within both one-band and three-band models\cite{astrid1,fwave}, depending on the doping level and the interaction parameters. However, there is no direct experimental evidence for a triplet state in cuprates so far. It is within this context that the recent discovery of Superconductivity (SC) in Ba$_2$CuO$_{3+\delta}$ (BCO)\cite{lietal} becomes particularly significant, as it may provide a new platform to explore the nature of the pairing symmetry in cuprates.   
	
	The Unit cell of Ba$_2$CuO$_{3.25}$ has two primary layers: layer-I and layer-II. Layer-I exhibits octahedral and square planar Cu-O complexes alternately stacked along the b-direction and layer-II has only square planar complexes along the b-axis, see Fig. \ref{ref:fig1}(a). This two layer feature originates from the missing oxygen atoms in Ba$_2$CuO$_{4}$ lattice. The oxygen k-edge x-ray absorption spectra\cite{lietal,xasbco_2} have provided evidence that points towards such a layered arrangement. Electronic structure calculations show that the interlayer hybridization has a pronounced effect on the effective band structure, leading to a shift in the Van Hove singularity (VHS) and $d_{x^2-y^2}$ orbital  moves below the Fermi level\cite{bco_dft1,our1}. However, as we increase the hole doping, the $d_{x^2-y^2}$ orbital crosses the Fermi level and contributes significantly to the VHS. This is crucial for determining the pairing symmetry in BCO. 
	
	In this paper, we investigate the variations in superconducting pairing symmetry and strength as a function of hole doping. We consider a 14-orbital basis, which is subsequently downfolded to a five-orbital basis\cite{our1}. This includes the $d_{z^2}$ and $d_{x^2-y^2}$ orbitals from the CuO$_6$ octahedra of the Cu(1) atom in layer-I, and the $d_{b^2-c^2}$ orbital from the chain state of the Cu(2) atom in layer-I and the Cu(3)/Cu(4) atoms in layer-II. This Hamiltonian accurately replicates the low-energy bands observed in density functional theory (DFT) calculations. Notably, due to octahedral compression, the Cu-$d_{x^2-y^2}$ orbital is positioned below the Fermi level, see \ref{ref:fig1}(b). As we increase hole doping, this orbital becomes crucial in determining the superconducting pairing symmetry. Our model assumes that superconducting pairing interactions arise from spin-fluctuation mechanisms, incorporating multiband Hubbard interactions within the framework of the weak coupling random phase approximation (RPA). The most robust solution of this pairing interaction indicates towards dominant superconducting gap. The dominant pairing strength in the singlet channel originates from a nesting vector of $\pm(\pi$-$\delta$,$\pi$-$\delta$). Consequently, the superconducting gap function changes sign between momentum vectors connected by this nesting vector, leading to a $d_{x^2-y^2}$ pairing symmetry.
	
	Prompted by the recent discovery of SC in BCO at high doping level (40\%), we aim to elucidate the superconducting pairing symmetry across the entire range of hole doping. We ascertain that below a critical doping threshold ($x_c$), the superconducting pairing symmetry predominantly exhibits the conventional $d_{x^2-y^2}$ type. However, with increased doping, the superconducting pairing symmetry transitions to a multi-band triplet solution. Notably, this feature demonstrates a significant robustness against Hund's coupling. This change in pairing symmetry we attribute to the inter-orbital nesting vectors that connect the $d_{z^2}$ and $d_{x^2-y^2}$ FSs. While such triplet symmetry has not yet been observed in Ba$_2$CuO$_{3.2}$, our findings suggest that through chemical doping or the application of pressure, this type of superconducting gap structure could be realized in future experiments.
	
	The rest of the paper is organised as follows: Section II we provide a summary of the tight-binding model, along with the mechanisms underlying the multiband RPA spin fluctuation approach. In Section III, we delve into the FS topology, examining the nesting profile, RPA spin susceptibility, and pairing symmetry. This section also explores the variation in pairing strength as a function of doping and investigates the density of states in the SC state. Finally in section IV we give a comprehensive discussion and conclusion of our findings.
	
	\section{Method and Model}
	
	\subsection{Tight-binding model}
	We consider a 14-orbital, tight-binding (TB) Hamiltonian that effectively replicates the low energy density functional theory (DFT) band structure \cite{our1}. %within the energy range of ($\pm 0.5 eV$)[see detail calculations in \cite{our1}].
 In our model, the four Cu atoms in the unit cell contribute five Cu-$d$ orbitals, while the adjacent oxygen  atoms provide nine $p$ orbitals. The orbital weights contribution to the DFT results shows that oxygen bands lie deep inside the Fermi level. Using L\"owdin downfolding procedure we integrate out the oxygen bands and obtain an effective five-band model Hamiltonian. A more comprehensive analysis of this contribution is available in the appendix. The Hamiltonian is expressed as follows,
	\begin{equation}
		H({\bf k}) = \sum_{\alpha\beta } \sum_{\substack{{\bf k} \\ \sigma\in(\uparrow,\downarrow)}}\Big[ \xi_{\alpha\beta}({\bf k}) + \mu_{\alpha} \delta_{\alpha\beta}\Big] c_{{\bf k},\alpha,\sigma}^{\dagger}c_{{\bf  k},\beta,\sigma}. \label{ho}
	\end{equation}
	$ \xi_{\alpha\beta}({\bf k})$ is the TB matrix element fitted with the DFT bands. Fermion creation and annihilation operator are denoted by  $c^{\dagger}_{{\bf k},\alpha,\sigma}$ and $c_{{\bf k},\beta, \sigma}$ respectively. The onsite energy for orbital $\alpha$  is $\mu_{\alpha}$
	
	\subsection{Multiband RPA susceptibility}
	We use the multi-band Hubbard model to study the topology of FS and corresponding spin-fluctuation potential. The Hamiltonian of the Hubbard model is given by \cite{graser},
	
\begin{eqnarray}
	H_{\rm int} &=& 	\sum_{\alpha,{\bf q}}U n_{\alpha \uparrow}({\bf q})  n_{\alpha \downarrow}(-{\bf q})+  \sum_{\alpha \ne \beta} \sum_{{\bf q}}  \frac{V }{2} n_{\alpha }({\bf q})  n_{\beta }(-{\bf q})  \nonumber\\
	&&-  \sum_{\alpha \ne \beta} \sum_{{\bf q}} \frac{J}{2} {\bf S}_{\alpha }({\bf q}) \cdot  {\bf S}_{\beta }(-{\bf q})  \label{Hint}
\end{eqnarray}

	Where $U$ and $V$ are the intra-orbital and inter-orbital Hubbard interaction between Cu-$d$ orbitals and $J_{H}$ is the Hund coupling. 
 
 %({\red{Is there a particular reason we wrote this in a more generic form than the spin SO(3) form that is being utilized in the calculation?}})
	
	 %N is the volume of the phase space. 
  Using perturbative expansion of the spin density and charge-density correlation function we obtain random-phase approximation (RPA) spin and charge susceptibilities,
	\begin{eqnarray}\label{spin_sus}
		\tilde{\chi}_{\rm s/c}({\bf q})= \tilde{\chi}_{0}({\bf q})\left(\tilde{\mathbb{I}} \mp \tilde{U}_{s/c}\tilde{\chi}_{0}({\bf q})\right)^{-1},
		\label{RPA}
	\end{eqnarray}
	
	The nonzero components of onsite Hubbard interactions for spin and charge fluctuation are $\tilde{U}_{s}$ and $\tilde{U}_{c}$\cite{fwave, graser}. The bare susceptibility is enhanced at the nesting wave vector which leads to a corresponding enhanced peak in the RPA spin susceptibility. The overall momentum space structure of the susceptibility can in general be more complex in multi orbital systems owing to the presence of matrix elements. In general, due to the presence of $(1-\tilde{U}\chi_{0})$ in the denominator for the RPA spin susceptibility, the contribution from the spin channel is enhanced compared to the charge channel that contains a $(1+\tilde{U}\chi_{0})$ contribution.
	
	\subsection{Superconducting pairing symmetry}
	
	Superconducting pairing in Cu -d electrons is mediated via spin fluctuations\cite{our1}.  We calculate the spin-fluctuation pairing potential by expanding the $ H_{\rm int}$ from Eq.~\eqref{Hint} into a perturbation series and collecting bubble and ladder diagrams. The effective Hamiltonian we obtain as \cite{fwave,nickelates1},
	\begin{eqnarray}
		H_{\rm eff} = \sum_{\alpha\beta\gamma\delta}\sum_{{\bf kq},\sigma\sigma'} \Gamma_{\alpha\beta}^{\gamma\delta}({\bf q}) c_{\alpha \sigma}^{\dagger}({\bf k})c_{\beta\sigma'}^{\dagger}(-{\bf  k})&&c_{\gamma\sigma'}({\bf -k-q}) \nonumber\\
		&& c_{\delta\sigma}({\bf k+q}).
		\label{Hintpair}
	\end{eqnarray}
	
	Here, the pairing potential is a tensor of four orbital indices. For singlet and triplet channels the spin-fluctuation pairing potential is given by \cite{SCrepulsive,SCcuprates,SCpnictides,SCHF,SCorganics,SCTMDC},
	\begin{subequations}
		\begin{eqnarray}
			\tilde{\Gamma}_{S}({\bf q})&=&\frac{1}{2}\big[3{\tilde U}_{s}{\tilde \chi}_{s}({\bf q}){\tilde U}_{s} - {\tilde U}_{c}{\tilde \chi}_{c}({\bf q}){\tilde U}_{c} + {\tilde U}_{s}+{\tilde U}_{c}\big],
			\label{singlet}\\
			\tilde{\Gamma}_{T}({\bf q})&=& -\frac{1}{2}\big[{\tilde U}_{s}{\tilde \chi}_{s}({\bf q}){\tilde U}_{s} + {\tilde U}_{c}{\tilde \chi}_{c}({\bf q}){\tilde U}_{c}\big].
			\label{triplet}
		\end{eqnarray}
	\end{subequations}

	Using a unitary transformation we obtain pairing potential in the band basis. 
	
	\begin{eqnarray}
		\tilde{\Gamma}_{\mu\nu}({\bf k,q})=\sum_{\alpha\beta\gamma\delta} \Gamma_{\alpha\beta}^{\gamma\delta}({\bf q})  \psi^{\mu\dagger}_{\alpha}({\bf k})\psi^{\mu\dagger}_{\beta}(-{\bf k}) &&\psi^{\nu}_{\gamma}({\bf -k-q}) \nonumber\\
		&&\psi^{\nu}_{\delta} ({\bf k+q}) \nonumber \\
	\end{eqnarray}
	Where $(\mu,\nu)$ represent the band indices, and $\psi^{\mu}_{\alpha}({\bf k})$ is the eigenvector component corresponding to orbital $\alpha$, band $\mu$ and calculated at the wave vector ${\bf k}$. We obtain superconducting pairing symmetry by solving the linearized gap equation,
	\begin{eqnarray}
		\Delta_{\mu}({\bf k})= -\lambda\frac{1}{N}\sum_{\nu,{\bf q}}\tilde{\Gamma}_{\mu\nu}({\bf k,q})\Delta_{\nu}({\bf k+q}).
		\label{SC2}
	\end{eqnarray}
	$\lambda$ is known as a superconducting coupling constant. By solving  Eq. \eqref{SC2} we obtain pairing eigenfunction for largest eigenvalue. This largest eigenvalue determines the stability of superconducting gap function $\Delta({\rm \bf k})$\cite{SCrepulsive}.    
	
The unconventional SC within spin fluctuation theory originates from a repulsive interaction that usually favours a sign change of the superconducting gap over the FS. Since $\chi_{s}$ (see  Eq. \eqref{RPA}) is positive and larger than $\chi_{c}$, pairing potential Eq. \eqref{singlet} is repulsive. The only possible solution of the gap equation for repulsive interaction is when $\Delta$ (see  Eq. \eqref{SC2} ) changes sign between momentum vectors ${\bf k}$ and ${\bf k + Q}$. This leads to an anisotropic solution of the gap function in the momentum space whose underlying symmetry transforms according to the irreducible representation of the crystals point group symmetry. 
	\section{Results}
	
	\subsection{Electronic structure}

				\begin{figure}
		\begin{center}
			\rotatebox{0}{\includegraphics[width=0.49\textwidth]{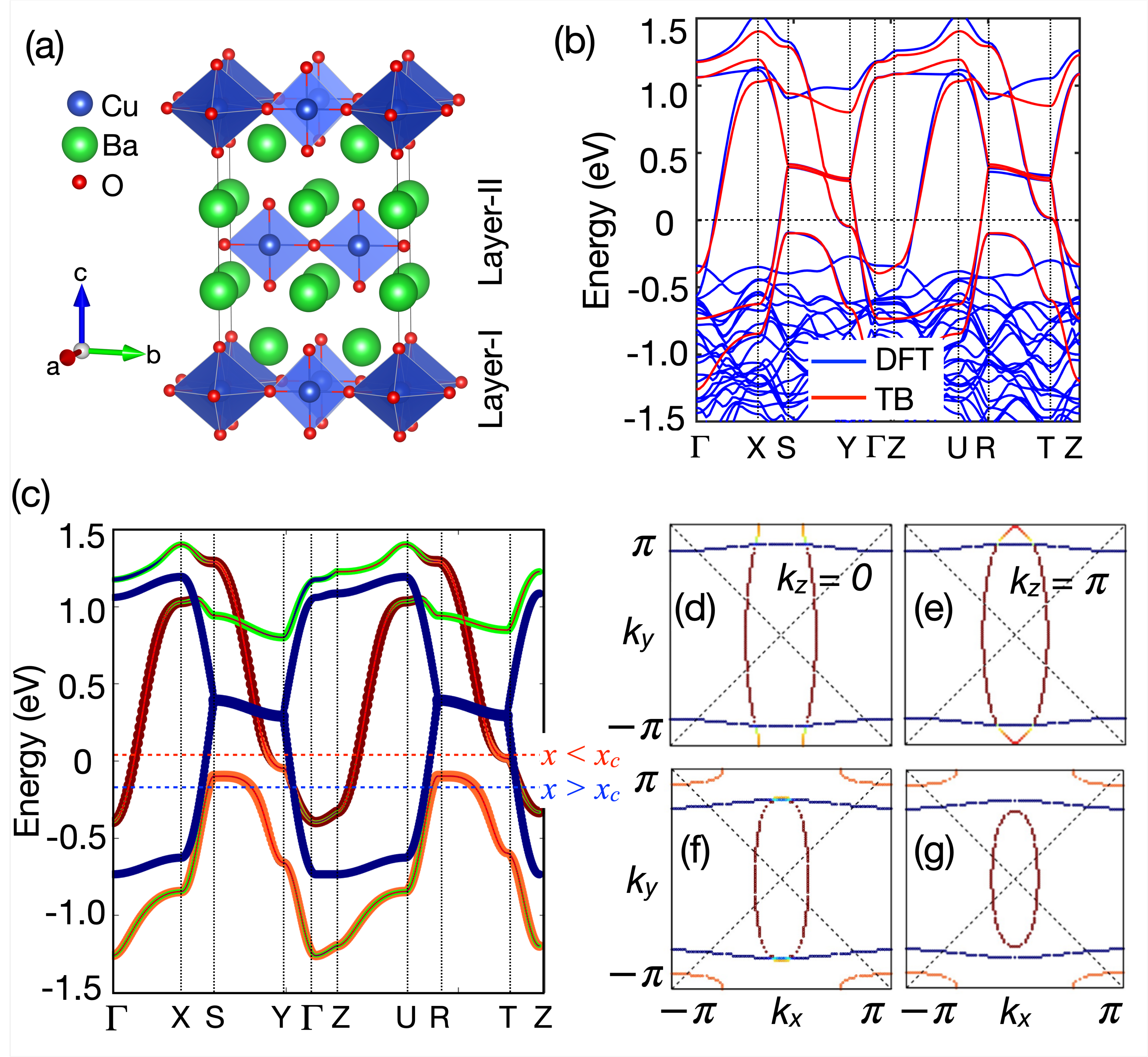}}
		\end{center}

  \caption{(a) Crystal structure of Ba$_2$CuO$_{3.25}$. Layer-I consists of copper in octahedral coordination (Cu(1)) and square planar complexes (Cu(2)), while Layer-II  features Cu(3)/Cu(4) in square planar complexes, aligned along the $b$-axis. (b) Tight binding model fitted with DFT band structure \cite{our1}. (c) Orbital resolved tight binding bands of BCO.  The red and blue dashed lines indicate two specific doping levels at $E_{F}$  of 0.0778 eV and -0.0778 eV, based on rigid band approximations. These correspond to doping levels of $x$ = 0.24 and $x$ = 0.37, respectively. The critical doping level, where SC pairing symmetry transitions from singlet to triplet, is shown in  Fig. \ref{ref:fig3}. The critical values are $x_c$=0.29, 0.33 for $J_{H}=0$ and $J_{H} \neq
  0$  respectively.  Fermi surfaces are shown at two different $k_{z}$ cuts (0 and $\pi$)  for two different doping levels, $x< x_c$ in (d)-(e) and   $x>x_c$ in (f)-(g) respectively.  The colour scheme for different orbitals are, $d_{z^2}$ (brick red), $d_{x^{2}-y^{2}}$ (orange), Cu(2)-$d_{b^{2}-c^{2}}$(green), Cu(3)/Cu(4)- $d_{b^{2}-c^{2}}$ (navy blue).     }
	%	\caption{ (a)-(b)Electronic structures of the downfolded  five-band noninteracting
		%	model, for two representative values of doping ($x$). (d)-(g) Corresponding FSs are shown for the same two cases presented
			%in the upper panel at two different $k_{z}$ cuts. }

		\label{ref:fig1}
	\end{figure}

In this study, we begin by analyzing the electronic structure and FS topology at two distinct doping levels. We choose two representative  doping levels, $x$=0.24 and $x$ = 0.37, corresponding to a Fermi energy ($E_F$) of 0.0778 eV and -0.0778 eV, respectively. The electronic band structures are depicted in Fig. \ref{ref:fig1}(c). The red and blue horizontal dashed line in \ref{ref:fig1}(c) denotes the low($x$=0.24) and high( $x$ = 0.37) doping values. Our findings reveal that increasing hole doping leads to a reduction in the electron filling of the $d_{x^2-y^2}$ orbital. This creates additional hole pockets reminiscent of cuprate superconductors\cite{fwave}. Notably, when the $d_{x^2-y^2}$ hole pocket becomes fully depleted, the FS is predominantly characterized by open electron-like pockets of $d_{z^2}$ at $k_z$=0 and closed electron pockets at $k_z$=$\pi$, with an enhanced contribution from $d_{b^2-c^2}$ orbitals on the open electron pockets persisting across the entire range of hole doping. Experimentally, the FS results can be corroborated by ARPES measurements conducted on BCO samples. Intriguingly, the presence of the hole pocket introduces additional nesting vectors, absent in the 40\% hole doping regime. In the following sections we will demonstrate that these alterations in the $d_{x^2-y^2}$ FS significantly impact the overall pairing potential and, consequently, the pairing symmetry within the BCO system.
		\subsection{Evolution of FS nesting with  doping}
				\begin{figure}
		\begin{center}
			\rotatebox{0}{\includegraphics[width=0.5\textwidth]{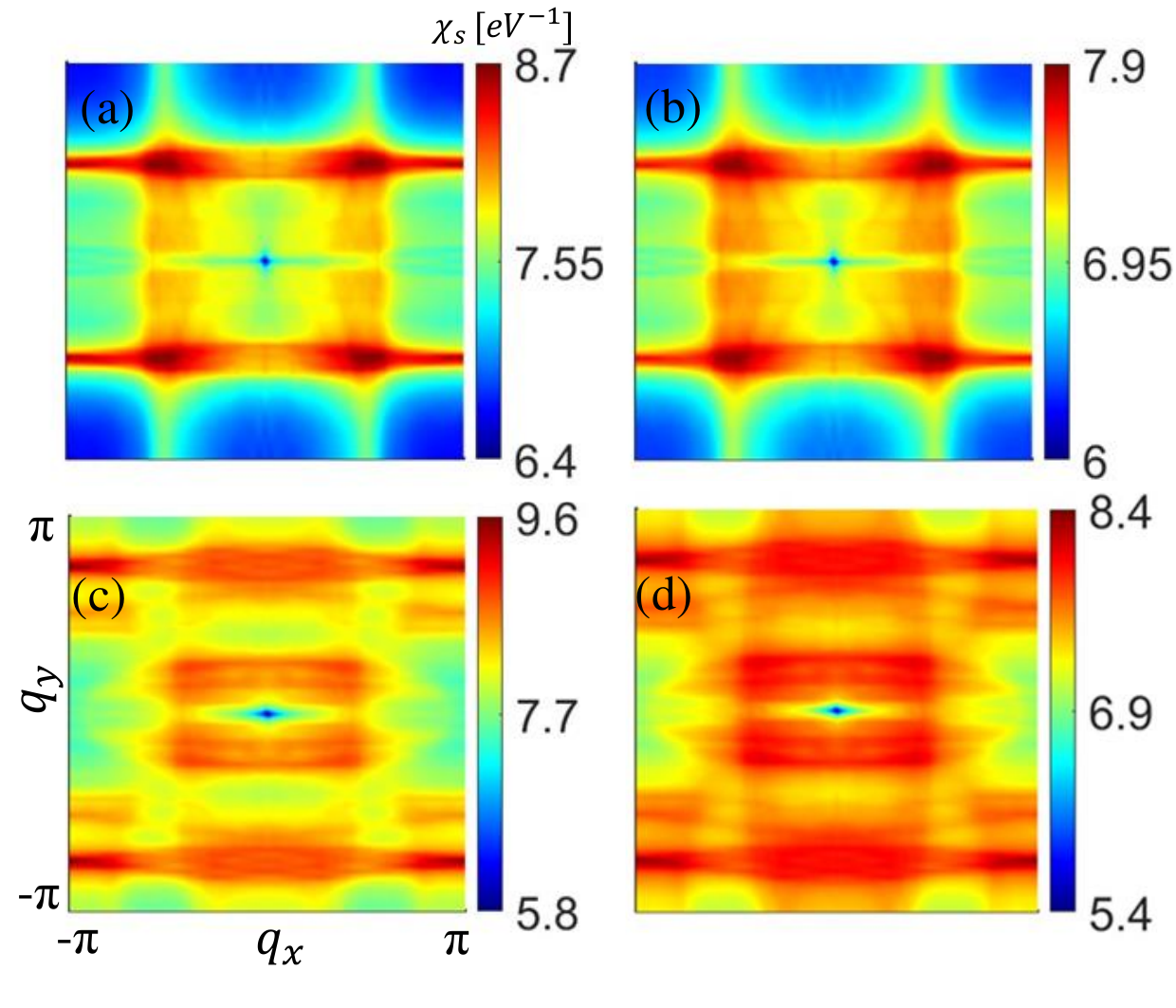}}
		\end{center}
		\caption{ (a)-(d) Physical RPA spin susceptibilities for two representation dopings. (a-b) is for $x< x_{c}$ with  $J_{H}$ = 0  and $\frac{U}{3}$ repectively. (c-d) is for $x> x_{c}$ with  $J_{H}$ = 0  and $\frac{U}{3}$ repectively. }
		\label{ref:fig2}
	\end{figure}	
 
\begin{figure}
		\begin{center}
			\rotatebox{0}{\includegraphics[width=0.5\textwidth]{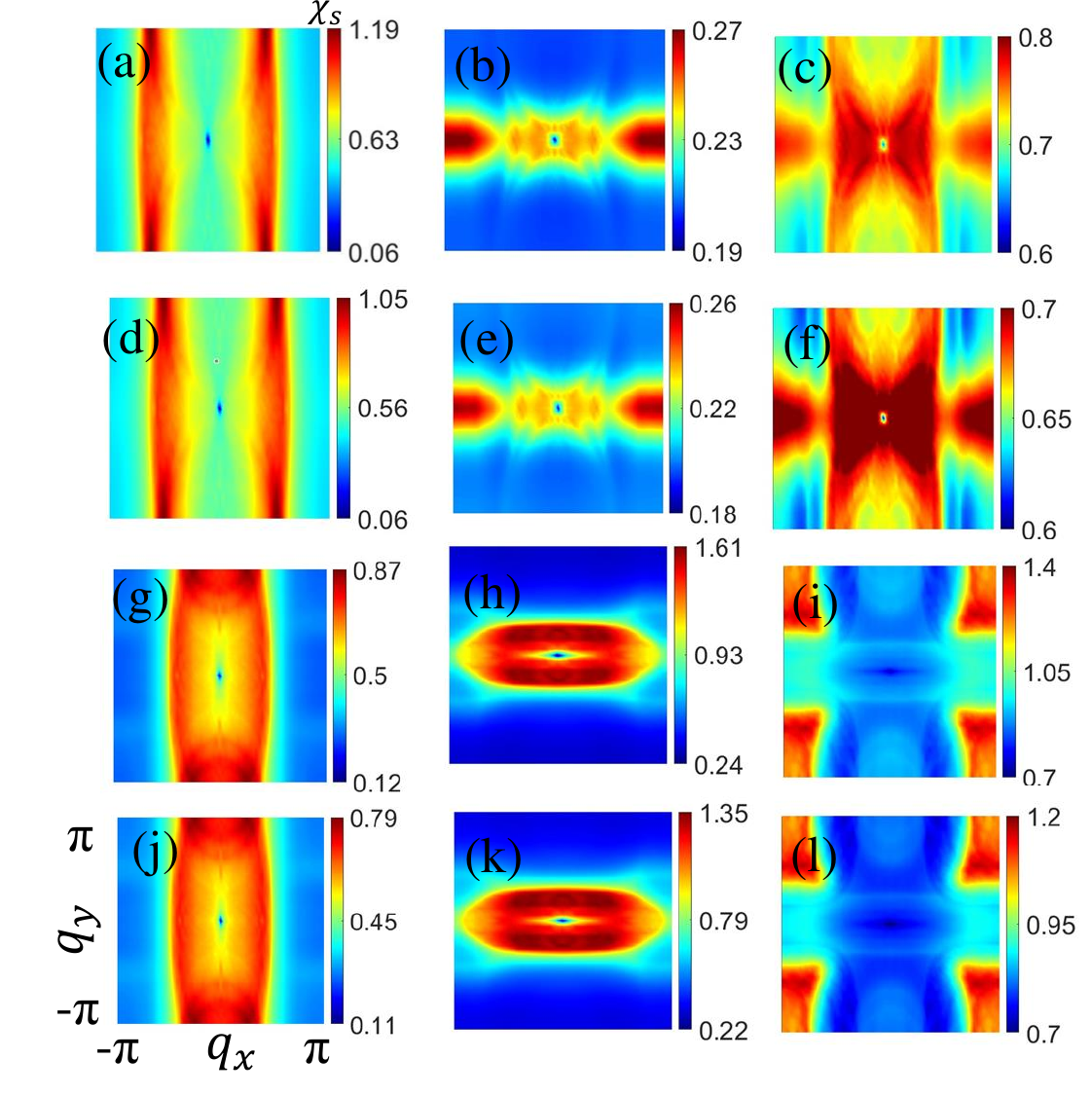}}
		\end{center}
		\caption{ (a)-(l) Orbital resolved RPA spin susceptibilities for two representation dopings. Column 1 and 2 belong to intra-orbital ( $\chi_{11}^{11}$,  $\chi_{22}^{22}$ ) and column 3 is for inter-orbital ($\chi_{12}^{12}$ ) contributions. Row 1 is for $x< x_{c}$ with  $J_{H}$ = 0 , row 2 for $x< x_{c}$ with  $J_{H} = U/3$,   Row 3 is for $x> x_{c}$ with  $J_{H}$ = 0 , row 4 for $x> x_{c}$ with  $J_{H} = U/3$.   }
		\label{ref:fig6}
	\end{figure}

	     In the following section, we delve into the evolution of the FS nesting profile as a function of hole doping concentration. For this purpose, we compute the RPA spin-susceptibility, as shown in Fig. \ref{ref:fig2} for $q_z$=0, and orbital-resolved components in Fig. \ref{ref:fig6}. At low doping, FS has a mixed character of $d_{z^2}$ and $d_{x^2-y^2}$ orbitals near (0, $\pi$) region. RPA susceptibility of the $d_{x^2-y^2}$ orbital is much lower than the inter-orbital contribution, as evident from Fig. \ref{ref:fig6}(b)-(c). We have found a pronounced peak at the physical spin susceptibility near the $\pm ( \pi-\delta$,$\pi-\delta$) wave-vector, in Fig. \ref{ref:fig2}, which we posit to be a precursor to a $d_{x^2-y^2}$-type superconducting gap.

	The FS displays a $k_z$ dependence, as depicted in  Fig. \ref{ref:fig1}(d)-(g). However, the distribution of orbital weight across the Fermi level is such that the RPA spin susceptibility exhibits negligible $q_z$ dependence. Consequently, we have focused solely on the $q_z$=0 components in our susceptibility calculations. With increasing hole doping, in addition to the smaller electron-like pocket at the center of the Brillouin zone (BZ), we have an additional hole pocket with $d_{x^2-y^2}$ character emerges near the corners of BZ [see Fig. \ref{ref:fig1}. This new hole pocket, reminiscent of typical cuprate superconductors which gives a nesting vector of ($\pi$, $\pi$) to the CuO$_4$ plane. However, in BCO, because of the structural anisotropy (relative to Ba$_2$CuO$_{4}$ ) the hole pocket significantly alters the nesting vector, as demonstrated in the susceptibility plot in Fig. \ref{ref:fig2}(c)-(d). Moreover, there is a notable reversal in the contributions of the $d_{x^2-y^2}$ and $d_{z^2}$ orbitals to the intra-orbital RPA susceptibility components in Fig. \ref{ref:fig6}(g)-(h). The contribution of the intraorbital $d_{x^2-y^2}$ is marginally more significant than the inter-orbital susceptibility, which in turn correlates with a weaker $d_{z^2}$ intraorbital susceptibility in Fig. \ref{ref:fig6}(g)-(i).
	
\begin{figure}
	\begin{center}
		\rotatebox{0}{\includegraphics[width=0.5\textwidth]{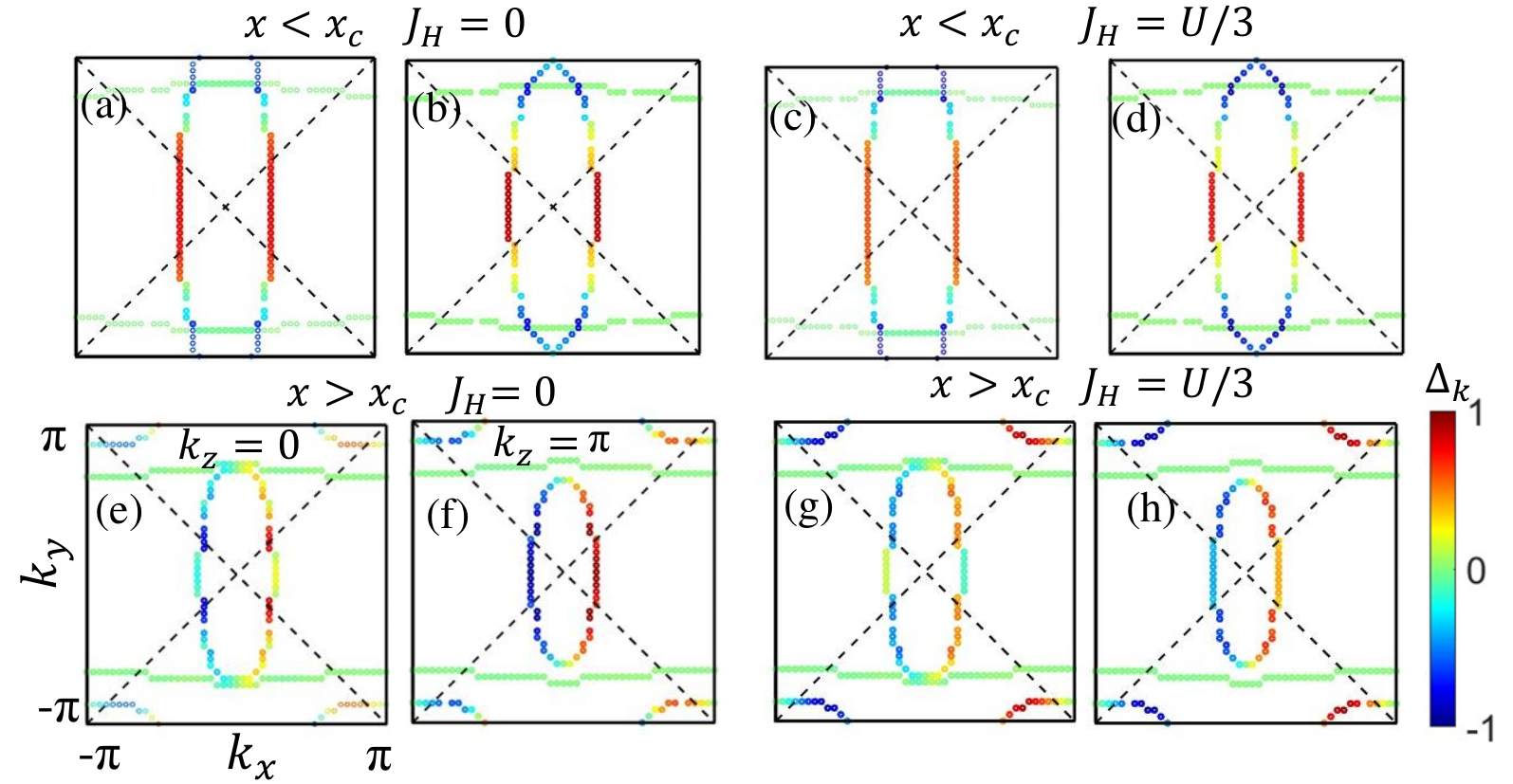}}
	\end{center}
	\caption{ (a)-(h)Computed pairing eigenfunction $\Delta$({\bf k}) for the leading eigenvalue, plotted on the corresponding FSs,
		for two representative values of doping ($x$). $x_{c}$ is the critical value of the doping at which leading SC gap changes from singlet to triplet pairing. }
	\label{ref:fig4}
\end{figure}

	Fig. \ref{ref:fig2}(b) and(d) demonstrate the influence of Hund's coupling in diminishing the effectiveness of charge screening, together with their orbital resolved components in Fig. \ref{ref:fig6}(d)-(f) and (j)-(l). It is noteworthy that in the absence of Hund's coupling, the peak value of the RPA spin susceptibility is larger, yet the overall nesting vector remains unchanged. Hence we expect that the nodal structure of the SC gap  does not change. Indeed, as will be demonstrated in Fig. \ref{ref:fig4}, the symmetry of the superconducting gap does not vary for a given doping level. However, since the superconducting pairing potential is directly proportional to the strength of the RPA spin susceptibility, the strength of the pairing symmetry is attenuated in the presence of Hund's coupling.
	
\subsection{Superconducting properties}

We now turn our attention to the doping dependence of SC in BCO. We include  doping levels comparable to those considered in our susceptibility discussions. In Fig. \ref{ref:fig4}, we present a plot of the superconducting gap on the FS, highlighting the largest pairing eigenvalue. The colorbar in this figure illustrates the sign change in the superconducting gap function. Two distinct solutions emerge at varying doping values. Our results indicate that the singlet potential solution of Eq.~\eqref{singlet} exhibits the highest pairing eigenvalue ($\lambda$) at lower doping levels. Conversely, at higher doping levels, the triplet channel, Eq.~\eqref{triplet}, gives the largest pairing strength.

As discussed earlier in the context of Fig. \ref{ref:fig2},  at low doping levels, the FS nesting at $(\pi-\delta, \pi-\delta)$ significantly contributes to the RPA spin susceptibility. The FS exhibits \(C_2\) symmetry, which is mirrored in the solution of the superconducting pairing potential. For doping levels below 40\%, the nesting condition within the CuO$_6$ octahedra fosters a pairing symmetry that fulfils the relation $\Delta(k+Q_1) = -\Delta(k)$ for the SC gap function, where $Q_1 = (\pi - \delta, \pi - \delta)$. At higher doping values $(x > x_c)$, the uniaxial nesting condition arising from the $d_{x^2-y^2}$ orbital yields a nesting vector $Q_2 = (\pi - \delta, q_y \approx \text{small})$, affecting all Fermi momentum vectors. In this scenario, the FS leads to a gap function with a $\sin(k_x)$ structure. This corresponds to the triplet solution that emerges when the hole pocket crosses the van Hove singularity point from below. Consequently, in the overdoped region of BCO, we observe a triplet $p$-wave solution.

\begin{figure}
		\begin{center}
			\rotatebox{0}{\includegraphics[width=0.5\textwidth]{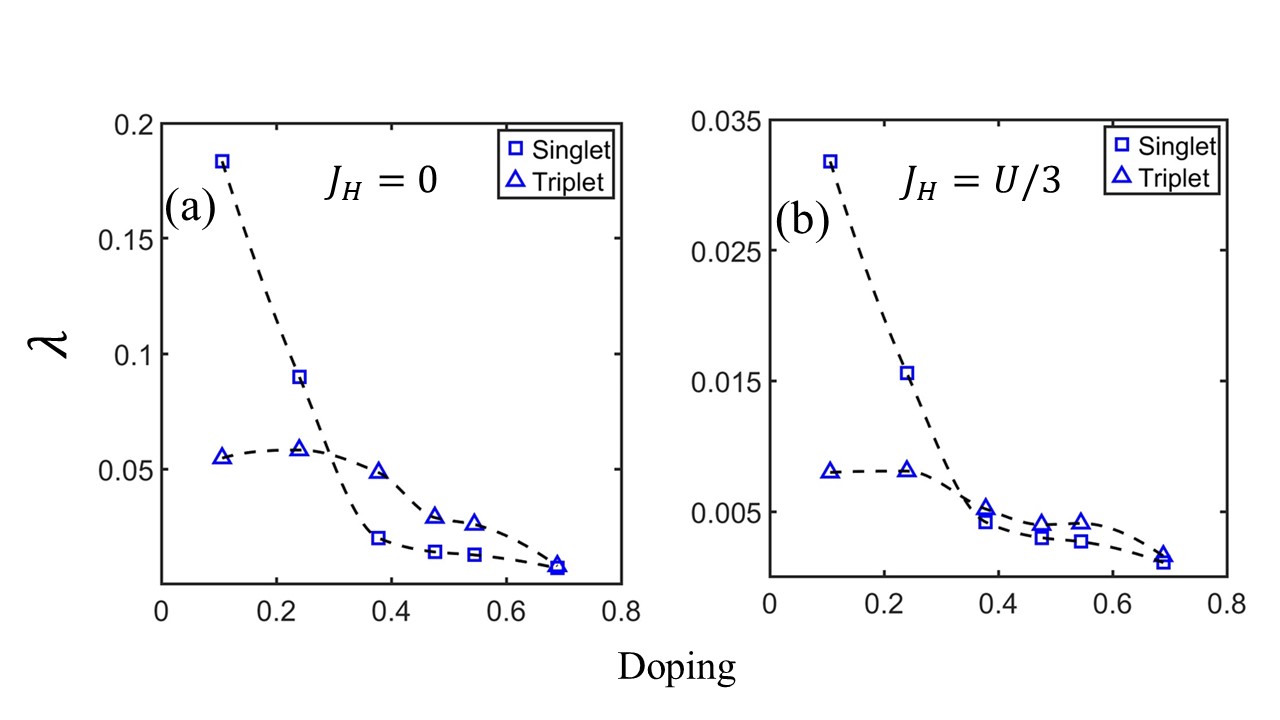}}
		\end{center}
		\caption{ (a)-(b) Doping-dependent SC coupling constant $\lambda$  for BCO for
			a choice of $J_{H}$ = 0 eV and $J_{H}=\frac{U}{3}$ eV. }
		\label{ref:fig3}
	\end{figure}

Next, we show the highest pairing eigenvalue across the entire hole-doped region, as depicted in Fig. \ref{ref:fig3}. The selection of interaction parameters is chosen from the Stoner criterion, which is essential for fulfilling the normal state paramagnetic solution. Superconducting pairing strength, $\lambda$, decreases with increased doping. This behaviour remains robust against Hund's coupling. Hence the doping dependence of $\lambda$ is fundamentally linked to the FS nesting properties. The observed decrease in $\lambda$ with hole doping is elucidated by examining the orbital-resolved DOS for BCO \cite{our1}. With hole doping, the DOS of the $d_{x^2-y^2}$ orbital increases, approaching the VHS, while the DOS of the $d_{z^2}$ orbital diminishes. Additionally, a comparative analysis of the FS volumes in two doping regions indicates an increase in FS volume with enhanced electron filling. This contributes significantly to the calculations of $\lambda$ using Eq. 7. 
The evolution of FS with doping is non monotonic, marked by the emergence of new orbitals ($d_{x^2-y^2}$). Within the framework of weak-coupling theory, orbital resolved SC gap is also expected to form in these new orbitals. %More sophisticated theoretical approaches, such as Fluctuation Exchange (FLEX) and Dynamical Mean Field Theory (DMFT), could more accurately capture these correlation effects, although they lie beyond the scope of the current study.

\subsection{Superconducting spectral function }

	We consider the Nambu spinor basis, $\Psi_{\bf k} =  \Big(\phi_{\bf k\sigma}, \phi^{\dagger}_{-\bf k-\sigma}\Big)^{T}$, to calculate the spectral function. Eigenstates of the non-interacting Hamiltonian($H({\bf k})$), see Eq.~\eqref{ho},  are denoted by $\phi_{\bf k\sigma}$. Using this, we construct mean-field Hamiltonian,
	
	\begin{eqnarray}
		H_{SC}= \begin{pmatrix}
			H({\bf k}) &	\tilde{\Delta}({\bf k})\\ 
			\tilde{\Delta}^{\dagger}(-{\bf k}) & -H^{\dagger}(-{\bf k}) 
		\end{pmatrix}.
	\label{Ham_sc}
	\end{eqnarray}
	
	Where $	\tilde{\Delta}({\bf k})$ is the SC gap functions we obtain from the self-consistent solution of Eq.~\eqref{SC2} for band $\nu$. We have adopted  tilde notation to indicate that, the SC gap exhibits multi-band components. To elucidate this aspect, the off-diagonal terms are explicitly written as follows,

%\begin{eqnarray}
%&&\Delta({\bf k})
%\nonumber \\\nonumber&=& \left(\begin{matrix}
%		& \Delta_{d_{z^2}}({\bf k}) & .  &. & .  &.   \\
%		&.  & \Delta_{d_{x^2 -y^2}}({\bf k})  &. & .  &.  \\
%		&.  & . &\Delta_{d_{b^2 -c^2}}({\bf k}) & .  &. \\
%		&.  & .  &. & \Delta_{d_{b^2 -c^2}}({\bf k})  &.  \\
%		&.  & .  &. & .  &\Delta_{d_{b^2 -c^2}}({\bf k}) \\
%		\end{matrix}\right). \nonumber\\ 
%	\label{Ham_spec}
%\end{eqnarray}

\begin{eqnarray}
	\tilde{\Delta}({\bf k})
	 = \left(\begin{matrix}
		& \Delta_{1}({\bf k}) & .  &. & .  &.   \\
		&.  & \Delta_{2}({\bf k})  &. & .  &.  \\
		&.  & . &\Delta_{3}({\bf k}) & .  &. \\
		&.  & .  &. & \Delta_{4}({\bf k})  &.  \\
		&.  & .  &. & .  &\Delta_{5}({\bf k}) \\
	\end{matrix}\right). \nonumber\\ 
	\label{Ham_spec}
\end{eqnarray}

\begin{figure}
	\begin{center}
		\rotatebox{0}{\includegraphics[width=0.5\textwidth]{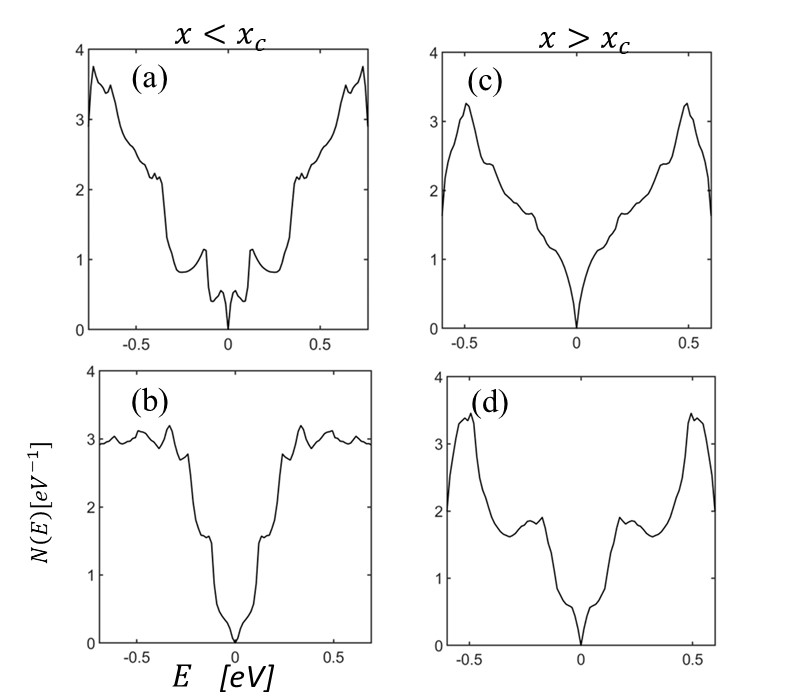}}
	\end{center}
	\caption{ (a)-(d) Total DOS for the SC state plotted for two representative values of doping ($x$).  (a-b) is for $x< x_{c}$ with  $J_{H}$ = 0  and $\frac{U}{3}$ repectively. (c-d) is for $x> x_{c}$ with  $J_{H}$ = 0  and $\frac{U}{3}$ repectively.  }
	\label{ref:fig5}
\end{figure}

The spectral functions we obtain from,  $\tilde{A}({\bf k},\omega) =  -\frac{1}{\pi} {\rm Im}\Big[\frac{1}{(\omega +i\delta) {{\bf {\it I}}} - H_{SC}}\Big]$. We have shown in Fig. \ref{ref:fig5}. the imginary part of the spectral function summed over all momentum.

	%	\subsection{Superconducting properties}

The SC gap function we obtain from the spin-fluctuation calculation at the Fermi momentum points. To extend these results across the entire BZ, we employ a Radial Basis Function (RBF)\cite{rbf1} for the extrapolation of two-dimensional data points for different $k_z$ values. Using this extrapolated gap function we obtain the spectral function and density of states.

In the low hole doping regime ($x < x_c$), the electron pockets at the Fermi level predominantly arise from the Cu(1) $d_{z^2}$ and Cu(3) $d_{b^2-c^2}$ orbitals in the singlet channel, whereas the Cu(1) $d_{x^2-y^2}$ and Cu(2)/Cu(4) planar atoms exhibit negligible contributions to superconductivity. Conversely, in the high hole doping regime ($x > x_c$), the emergence of an additional hole pocket attributable to the Cu(1) $d_{x^2-y^2}$ orbital facilitates the triplet channel. The SC gap structure, as previously discussed, varies significantly between these two channels. In Fig. \ref{ref:fig5} we  illustrates DOS in the SC state, highlighting the contributions from the Cu(1) $d_{z^2}$ and Cu(3) $d_{b^2-c^2}$ orbitals in (a) and (b) respectively, and an additional Cu(1) $d_{x^2-y^2}$ orbital contribution in (c). In Fig. \ref{ref:fig5}(b), (d)  we have shown DOS when Hund’s coupling is present.  

In presence of Hund’s coupling the V-shape feature is prominent. Otherwise DOS have a mixed V and U shape character. This DOS feature's robustness against variations in Hund's coupling is significant. However, the DOS differences between singlet and triplet states are primarily attributed to the multi-band structure of the SC-gap function. The observed V-shaped feature suggests a nodal gap structure. The residual DOS is attributable to the SC originating predominantly from two (or three) orbitals in the singlet (or triplet) channel, additionally the hopping parameter substantially larger than the SC gap's maximum across the BZ. The SC gap diminishes exponentially faster towards the BZ boundary compared to the hopping parameter. Our findings offer a theoretical framework that can be empirically validated through scanning tunneling microscopy (STM) measurements on BCO at different doping conditions.

%$\phi_{\bf k\sigma}= \Big(c_{1\bf k\uparrow}, c_{2\bf k\uparrow} c_{3\bf k\uparrow}, c_{4\bf k\uparrow} c_{5\bf k\uparrow}\Big)^{T}$

%$Q_2$ = ($\pi$ - $\delta$, $q_y$ $\approx$ {\text small})

	\section{Discussions and Conclusions}

We have identified a triplet p-wave pairing in BCO, occurring at hole doping levels significantly above the conventional optimal doping range for cuprate superconductors. 	This contrasts with other cuprates, where the Cu-$d_{x^2-y^2}$ orbital typically leads to d-wave pairing due to the ($\pi,\pi$) nesting vector. The key difference in BCO lies in the octahedral compression, which suppresses the $d_{x^2-y^2}$  orbital relative to the $d_{z^2}$ orbital.  The absence of $d_{x^2-y^2}$ changes the AFM nesting vector different than $(\pi,\pi)$. The nesting vector is changed by shifts in the Fermi level, enabling the emergence of $d_{x^2-y^2}$ SC at lower doping levels. The absence of the $d_{x^2-y^2}$  orbital due to octahedral compression can be compensated by shifting above the VHS point through a rigid band shift. This adjustment introduces an additional hole pocket, which in turn suppresses the RPA susceptibility of the $d_{z^2}$ orbital, previously dominant in the absence of  $d_{x^2-y^2}$. Since the spin-fluctuation pairing potential is directly proportional to the RPA spin susceptibility, there is now a substantial contribution at the $Q_2 = (\pi - \delta, q_y \approx \text{small})$ nesting vector. Consequently, the SC gap function exhibits a change in sign corresponding to the reversal of the momentum vector.
	
Triplet odd parity SC exhibits a range of fascinating applications, including in superconductor/ferromagnet heterostructures and topological SC, leading to the Majorana modes\cite{majorana1}, etc. Notably, triplet odd parity SC has been recently observed in the heavy fermion superconductor CeRh$_{2}$As$_{2}$\cite{tripletsc1}. Additionally, a handful of studies on cuprates have indicated the presence of p-wave SC \cite{TSC,DasNodelessSC,Gupta}. However similar results in cuprates is still lacking. In this context, our results could serve as a motivation for experimental investigations into the signatures of triplet SC in doped BCO. Such investigations might include field angle magnetic field measurements, nuclear magnetic resonance (NMR), Knight shift measurements, and Angle-Resolved Photoemission Spectroscopy (ARPES).	
	
	%\section{Appendix}
	\appendix
		\section{Details of the tight-binding Hamiltonian }\label{rpasus}
	
	The matrix representation of the SK-TB model Hamiltonian of BCO3.25 is shown in Eq. 2 of the main text. Here, the sub-matrix $H$ of the layer-I in basis set order $d_{z^2}$, $d_{x^2-y^2}$, $d_{x^2-y^2}$, $p_x$, $p_z$, $p_z$, $p_y$, and $p_y$ is:
	\begin{equation}
		H_{l1} = \left(\begin{array}{cccccccc}
			\xi_{1} &  \xi_{1,2} &  \xi_{1,3} & \xi_{1,4} & \xi_{1,5} & 0 &  \xi_{1,7} & 0 \\ \cline{1-1}
			\bord & \xi_{2} &  \xi_{2,3} &  \xi_{2,4} & \xi_{2,5} & \xi_{2,5}^{*} & 0 & 0 \\ \cline{2-2}
			&\bord  &  \xi_{3} & 0 & \xi_{2,5}^{*} & \xi_{2,5} & 0 & 0 \\ \cline{3-3}
			& &\bord & \mu^{(3)} & 0 & 0 & 0 & 0 \\  \cline{4-4}
			& {\rm h.c.} & &\bord & \mu^{(1)} & 0 & 0 & 0 \\  \cline{5-5}
			&  & & & \bord & \mu^{(1)} & 0 & 0 \\  \cline{6-6}
			&  & & & &\bord & \mu^{(1)} & 0 \\  \cline{7-7}
			&  & & & & &\bord & \mu^{(1)} \\  \cline{8-8}
		\end{array} \right)
	\end{equation}
	h.c. denotes the Hermitian conjugate of the upper-triangular matrix.
	
	The Hamiltonian sub-matrix for layer-II in orbital basis set of $d_{x^2-y^2}$, $d_{x^2-y^2}$, $p_z$, $p_z$, $p_y$ and $p_y$ is given as
	\begin{equation}
		H_{l2} = \left(
		\begin{array}{cccccc}
			\xi_{4} & \xi_{4,5} & \xi_{4,6} & 0 & 0 & 0 \\ \cline{1-1}
			\bord & \xi_{5} & 0 & \xi_{4,6} & 0 & 0 \\ \cline{2-2}
			&\bord & \mu^{(7)} & 0 & 0 & 0 \\ \cline{3-3}
			& {\rm h.c.} &\bord & \mu^{(7)} & 0 & 0 \\ \cline{4-4}
			& & &\bord & \mu^{(7)} & 0 \\ \cline{5-5}
			& & & &\bord & \mu^{(7)} \\ \cline{6-6}
		\end{array} \right)
	\end{equation}
	
	Further, the Hamiltonian sub-matrix containing the interaction between layer-I and layer-II is
	\begin{equation}
		H_{l1-l2} = \left(
		\begin{array}{cccccc}
			0 & 0 & 0 & 0 & \xi_{2}^{12} & -(\xi_{2}^{12})^{*} \\
			0 & 0 & 0 & 0 & 0 & 0 \\
			\xi_{1}^{12} & (\xi_{1}^{12})^{*} & 0 & 0 & \xi_{3}^{12} & -(\xi_{3}^{12})^{*} \\
			0 & 0 & 0 & 0 & 0 & 0 \\
			: &&&:&& : \\
			0 & 0 & 0 & 0 & 0 & 0 \\
		\end{array} \right)
	\end{equation}

	The components of the Hamiltonian matrices are found to be,
	
	\begin{eqnarray}
		\xi_{1}&=&t^{(5)}\cos(k_x) + \mu^{(4)} + t^{(15)}\cos(k_y)- 0.1\cos(2k_x)\\
		\xi_{2}&=&t^{(10)}\cos(ky) + t^{(11)}\cos(k_x)+ \mu^{(2)}\\
		\xi_{3}&=&\xi_{2}+\mu^{(5)}\\
		\xi_{4} &=&t^{(14)}\cos(k_x)+\mu^{(6)}\\
		\xi_{5}&=&\xi_{4}+0.02\\
		%		\xi_{6,6}&=&\mu^{(3)}\\
		%		\xi_{7,7}&=&\xi_{8,8}=\xi_{9,9}=\xi_{10,10}=\mu^{(1)}\\ 
		%		\xi_{11,11}&=&\xi_{12,12}=\xi_{13,13}=\xi_{14,14}=\mu^{(7)}\\ 
		\xi_{1,2}&=&t^{(3)}\cos(k_x)+t^{(9)}\cos(k_y)\\
		\xi_{1,3}&=&t^{(4)}\exp(ik_y/4)\\
		\xi_{2,3}&=&t^{(1)}\cos(k_y/2)\\
		%\xi_{1,4}&=&2t_{12}^{(1)}\exp(-iky/4)\cos(k_z/2)\cos(k_x/2)\\
		%	\xi_{1,5}&=&\xi_{1,4}\\
		%	\xi_{3,5}&=&\xi_{3,4}^{*}\\
		\xi_{1,4}&=&it^{(8)}\sin(k_x/2)\\
		\xi_{1,5}&=&it^{(8)}\sin(k_y/4)\\
		\xi_{1,7}&=&it^{(6)}\sin(0.156k_z)\\
		\xi_{2,4}&=&it^{(7)}\sin(k_x/2)\\
		\xi_{2,5}&=&t^{(4)}\exp(ik_y/4)\\
		%	\xi_{2,8}&=&\xi_{2,7}^{*}\\
		%	\xi_{3,7}&=&\xi_{2,8}\\
		%	\xi_{3,8}&=&\xi_{2,7}\\
		%	\xi_{1,14}&=&-\xi_{1,13}^{*}\\
		%	\xi_{3,14}&=&-\xi_{3,13}^{*}\\
		\xi_{4,5}&=&t^{(12)}\cos(k_y/2)\\
		\xi_{4,6}&=&it^{(13)}\sin(k_y/4)\\
		\xi_{1}^{12}&=&2t_{12}^{(1)}\exp(-ik_y/4)\cos(k_z/2)\cos(k_x/2)\\
		\xi_{2}^{12}&=&it_{12}^{(2)}\exp(ik_y/4)\sin(k_z/3)\cos(kx/2)\\
		\xi_{3}^{12}&=&it_{12}^{(2)}\exp(-ik_y/4)\sin(k_z/3)\cos(k_x/2)
		%	\xi_{5,12}&=&\xi_{4,11}\\
	\end{eqnarray}
	
	The tight-binding parameters are, $t^{(1-14)}$ = $\big[$0.9775, 0.0237, 0.3396, 0.005, -0.78, 0.2076, -0.46 , 0.0, 0.035, -0.119, -0.074, 0.91, 0.6928, -0.054, -0.06.$\big]$
	$\mu^{(1-7)}$ = $\big[$ -0.74, -0.31, -1.24, 0.5, 1.18, 0.28, -1.64 $\big]$
	$t_{12}^{(1-2)}$ = $\big[$ -0.02, -0.3$\big]$
	The L\"owdin method used for the downfolding mechanism can be explained by,
	\begin{equation}
		H^{\rm downfold}_{\alpha,\beta} = H_{\alpha,\beta} + \sum_{\gamma \neq \alpha}^{\prime} \frac{H_{\alpha,\gamma}(H_{\beta, \gamma})^*}{ H_{\alpha,\alpha} -H_{\gamma,\gamma}} 
	\end{equation}
	Here, $H^{\rm downfold}_{\alpha,\beta}$ is final 5$\times$5 downfolded Hamiltonian matrix, $\gamma$ contains O-$p$ -- O-$p$, Cu-$d$ -- O-$p$ orbitals interaction respectively which are projected on A. 
	%\Appendix	


\newpage
	
		\begin{thebibliography}{1}
					\bibitem{lietal}W. M. Li, J. F. Zhao, L. P. Cao, Z. Hu, Q. Z. Huang, X. C. Wang, Y. Liu, G. Q. Zhao, J. Zhang, Q. Q. Liu, R. Z. Yu, Y. W. Long, H. Wu, H. J. Lin, C. T. Chen, Z. Li, Z. Z. Gong, Z. Guguchia, J. S. Kim, G. R. Stewart, Y. J. Uemura, S. Uchida, and C. Q. Jin, PNAS {\bf 116}, 12156 (2019).
				      \bibitem{TsueiRMP}C. C. Tsuei and J. R. Kirtley,  Rev. Mod. Phys. {\bf 72}, 969 (2000).
			\bibitem{ScalapinoDwave}D. J. Scalapino, Phys. Rep. {\bf 250}, 329 (1995).
			\bibitem{KirtleyDwave}JR Kirtley, CC Tsuei, JZ Sun, CC Chi, Lock See Yu-Jahnes, A Gupta, M Rupp, MB Ketchen, Nature {\bf 373}, 225-228 (1995).
			
			\bibitem{wu} D. H. Wu {\it {et. al.,}} Phys. Rev. Lett. {\bf{70}}, 85
			(1993); S. M. Anlage {\it {et. al.,}} Phys. Rev. B. {\bf{50}}, 523
			(1994).
			
			\bibitem{andreone} A. Andreone {\it {et. al.,}} Phys. Rev. B.
			{\bf{49}}, 6392 (1994).
			\bibitem{schneider} C.W. Schneider, Z.H. Barber, J.E. Evetts, S.N. Mao, X.X. Xi, T. Venkatesan
			Physica C, {233}, 77 (1994).
			 \bibitem{LSCO2000}A. Ino, C. Kim, M. Nakamura, T. Yoshida, T. Mizokawa, Z.-X. Shen, A. Fujimori, T. Kakeshita, H. Eisaki, and S. Uchida, 
			%Electronic structure of La$_{2-x}$Sr$_x$CuO$_4$ in the vicinity of the superconductor-insulator transition. 
			Phys. Rev. B {\bf 62}, 4137 (2000).
			
			\bibitem{LSCO2013}E. Razzoli, G. Drachuck, A. Keren, M. Radovic, N. C. Plumb, J. Chang, Y.-B. Huang, H. Ding, J. Mesot, and M. Shi, 
			%Evolution from a nodeless gap to $d_{x^2Ã¢Ë†â€™y^2}$-wave in underdoped La$_{2-x}$Sr$_x$CuO$_4$. 
			Phys. Rev. Lett. {\bf 110}, 047004 (2013).
			
			\bibitem{bi22122006}K. Tanaka, W. S. Lee, D. H. Lu, A. Fujimori, T. Fujii, Risdiana, I. Terasaki, D. J. Scalapino, T. P. Devereaux, Z. Hussain, and Z.-X. Shen, 
			%Distinct Fermi-momentum-dependent energy gaps in deeply underdoped Bi2212. 
			Science {\bf 314}, 1910 (2006).
			
			\bibitem{bi2212} I. M. Vishik, {\it et al.} 
			%Phase competition in trisected superconducting dome. 
			Proc. Nat. Acad. Sci. (USA) {\bf 109}, 18332 (2012).
			
			\bibitem{bi2201} Y. Peng, J. Meng, D.Mou, J. He, L. Zhao, Y.Wu, G. Liu, X. Dong, S. He, J. Zhang, X. Wang, Q. Peng, Z. Wang, S. Zhang, F. Yang, C. Chen, Z. Xu, T. K. Lee, X. J. Zhou, 
			%Disappearance of nodal gap across the insulatorÃ¢â‚¬â€œsuperconductor transition in a copper-oxide superconductor. 
			Nat. Commun. {\bf 4}, 2459 (2013).
			
			\bibitem{ccoc} K. M. Shen, {\it et al.}, 
			%Fully gapped single-particle excitations in lightly doped cuprates. 
			Phys. Rev. B {\bf 69}, 054503 (2004).
			
			\bibitem{YBCO}D. Gustafsson, D. Golubev, M. Fogelström, T. Claeson, S. Kubatkin, T. Bauch, and F. Lombardi, 
			%Fully gapped superconductivity in a nanometre-size YBa$_2$Cu$_3$O$_{7Ã¢â‚¬â€œ\delta}$ island enhanced by a magnetic field. 
			Nat. Nanotechnology {\bf 8}, 25-30 (2013).
			
			\bibitem{scalapino_rmp1}D. J. Scalapino, Rev. Mod. Phys. {\bf 84}, 1383 (2012).
			
			\bibitem{astrid1}A. T. Rømer, A. Kreisel, I. Eremin, M. A. Malakhov, T. A. Maier, P. J. Hirschfeld, and B. M. Andersen, Phys. Rev. B {\bf 92}, 104505  (2015).
		
		\bibitem{fwave} Priyo Adhikary and Tanmoy Das, Phys. Rev. B {\bf 101}, 214517 (2020).
		

		
		\bibitem{xasbco_2}R. Fumagalli, A. Nag, S. Agrestini, M. Garcia-Fernandez, A. C.
		Walters, D. Betto, N. B. Brookes, L. Braicovich, K.-J. Zhou, G.
		Ghiringhelli, and M.Moretti Sala, Physica C: Supercond. Appl.
		{\bf 581}, 1353810 (2021).
				\bibitem{bco_dft1}Paul Worm, Motoharu Kitatani, Jan M. Tomczak, Liang Si, and Karsten Held, Phys. Rev. B {\bf 105}, 085110 (2022).
				\bibitem{our1}Priyo Adhikary, Mayank Gupta, Amit Chauhan, Sashi Satpathy, Shantanu Mukherjee, B. R. K. Nanda,  	arXiv:2310.05603.
		
		\bibitem{hightc_1}Q. Q. Liu, H. Yang, X. M. Qin, Y. Yu, L. X. Yang, F. Y. Li, R. C. Yu, C. Q. Jin, and S. Uchida, Phys. Rev. B {\bf 74}, 100506(R) (2006).
		  \bibitem{graser}S. Graser, T. A. Maier, P. J. Hirschfeld, D. J. Scalapino, New J. Phys. {\bf 11}, 025016 (2009).
		  		
		  
		  \bibitem{nickelates1} Priyo Adhikary, Subhadeep Bandyopadhyay, Tanmoy Das, Indra Dasgupta, and Tanusri Saha-Dasgupta
		  Phys. Rev. B {\bf 102}, 100501(R) (2020).
		  	
		  \bibitem{SCcuprates} D. J. Scalapino, E. Loh, Jr., and J. E. Hirsch, Phys. Rev. B {\bf 34}, 8190 (R) (1986); {\it ibid} Phys. Rev. B {\bf 34}, 6420 (1986); J. R. Schrieffer, {\it Theory of Superconductivity} (W. A. Benjamin, New York)  (1964); J. R. Schrieffer, X. G. Wen, and S. C. Zhang, Phys. Rev. B {\bf 39}, 11663 (1989); P. Monthoux, A. V. Balatsky, and D. Pines, Phys. Rev. Lett. {\bf 67}, 3448 (1991); M. Sigrist, and Kazuo Ueda, Rev. Mod. Phys. {\bf 63}, 239 (1991); J. C. Seamus Davis and Dung-Hai Lee, PNAS {\bf 110}, 17623-17630 (2013); T Das, RS Markiewicz, A Bansil, Adv. Phys. {\bf 63}, 151 (2014).
		  
		  \bibitem{SCrepulsive}D. J. Scalapino, Rev. Mod. Phys. {\bf 84}, 1383 (2012); A. V. Chubukov, D. Pines, J. Schmalian, In: Bennemann K.H., Ketterson J.B. (eds) The Physics of Superconductors. Springer, Berlin, Heidelberg; %DOI: https://doi.org/10.1007/978-3-642-55675-3_7; J. C. Seamus Davis and Dung-Hai Lee, PNAS {\bf 110}, 17623-17630 (2013); T Das, RS Markiewicz, A Bansil, Adv. Phys. {\bf 63}, 151 (2014).
		  \bibitem{SCpnictides}I. I. Mazin, D. J. Singh, M. D. Johannes, and M. H. Du, Phys. Rev. Lett. {\bf 101}, 057003S (2008); S. Graser, T. A. Maier, P. J. Hirschfeld, D. J. Scalapino, New J. Phys. {\bf 11}, 025016 (2009); Zi-Jian Yao, Jian-Xin Li, and Z D Wang, New J. Phys. {\bf 11}, 025009 (2009); T. Das, A. V. Balatsky, Phys. Rev. B {\bf 84}, 014521 (2011); A. Chubukov, Ann. Rev. Conden. Mat. Phys. {\bf 3}, 57-92 (2012).
		  
		  
		  
		  \bibitem{SCHF}Tetsuya Takimoto, Takashi Hotta, and Kazuo Ueda, Phys. Rev. B {\bf 69}, 104504 (2004); K. Kubo, Phys. Rev. B {\bf 69}, 104504 (2004); T. Das, J.-X. Zhu, M. J. Graf, Sci. Rep. {\bf 5,} 8632 (2015); Hiroaki Ikeda, Michi-To Suzuki, Ryotaro Arita, Phys. Rev. Lett. {\bf 114}, 147003 (2015); T. Nomoto, H. Ikeda, Phys. Rev. Lett. {\bf 117}, 217002 (2016); T. Nomoto, H. Ikeda, J. Phys. Soc. Jpn. {\bf 86}, 023703 (2017).	
		  
		  \bibitem{SCorganics}J. Schmalian, Phys. Rev. Lett. {\bf 81}, 4232 (1998); G. Saito, and Y. Yoshida, Chem Rec {\bf 11}, 124-145 (2011).
		  
		  \bibitem{SCTMDC}T. Das, and K. Dolui, Phys. Rev. B {\bf 91}, 094510 (2015); A. Bhattacharyya, et al., Phys. Rev. Lett. {\bf 122}, 147001 (2019).
		  
			  
	      \bibitem{rbf1}{  \url{ https://people.sc.fsu.edu/~jburkardt/m_src/rbf_interp_2d/rbf_interp_2d.html.}}
	      
	      \bibitem{majorana1} Y.F. Lv, {\it et al.} , Sci. Bulletin {\bf 62}, 852 (2017).
	      \bibitem{tripletsc1} J. F. Landaeta, and K. Dolui, Phys. Rev. X {\bf 12}, 031001 (2022);

      \bibitem{TSC}Y.-M. Lu, T. Xiang, and D.-H. Lee, 
      %Can deeply underdoped superconducting cuprates be topological superconductors? 
      Nat. Phys., {\bf 10}, 634–637(2014).
      
      \bibitem{DasNodelessSC}T. Das, arXiv:1312.0544.
      \bibitem{Gupta}Amit Gupta, and Debanand Sa, Eur. Phys. J. B {\bf 89}, 24 (2016).
	      
\end{thebibliography}
\end{document}